\documentclass[12pt]{article}
\usepackage{amssymb,amsmath,amsfonts}
\usepackage{graphicx}
\graphicspath{{Figures/}}
\usepackage{color}
\numberwithin{equation}{section}

\newcommand{\be}{\begin{equation}}
	\newcommand{\ee}{\end{equation}}
\newcommand{\bea}{\begin{eqnarray}}
	\newcommand{\eea}{\end{eqnarray}}

\newcommand{\e}{{\rm e}}

\renewcommand{\d}{{\rm d}}
\renewcommand{\i}{{\rm i}}

\newcommand{\grintl}{[\kern-.18em [}
\newcommand{\grintr}{]\kern-.18em ]}

\newcounter{resultcounter}[section]

\newtheorem{thm}[resultcounter]{Theorem}
\newtheorem{lem}[resultcounter]{Lemma}

\newtheorem{definition}[resultcounter]{Definition}

\def\bed{\begin{definition}}
	\def\eed{\end{definition}}
\def\one{{\mathchoice {\rm 1\mskip-4mu l} {\rm 1\mskip-4mu l} {\rm 1\mskip-4.5mu l} {\rm 1\mskip-5mu l}}}

\newcommand{\bbbone}{\mathchoice {\rm 1\mskip-4mu l} {\rm 1\mskip-4mu l}
	{\rm 1\mskip-4.5mu l} {\rm 1\mskip-5mu l}}

\newcommand{\s}{{\rm S}}
\renewcommand{\r}{{\rm R}}

\numberwithin{corollary}{section}

\numberwithin{remark}{section}
\newcommand{\mathsym}[1]{{}}
\newcommand{\unicode}[1]{{}}

\begin{document}
	
\title{Mean field dynamics  of\\
	 some open quantum systems
	}
	
	\author{\ \\
		 Marco Merkli\footnote{merkli@mun.ca,\ http://www.math.mun.ca/$\sim$merkli/} \qquad   Alireza Rafiyi\footnote{a.rafiyi@mun.ca}\\
		 \ \\
		 Department of Mathematics and Statistics \\
		Memorial University of Newfoundland\\
		St. John's, NL\\
		Canada\   A1C 5S7
		\ \\
	}

	\maketitle
	
\begin{abstract}
We consider a large number $N$ of quantum particles coupled via a mean field interaction to another quantum system (reservoir). Our main result is an expansion for the averages of observables, both of the particles and of the reservoir, in inverse powers of $\sqrt{N}$. The analysis is based directly on the Dyson series expansion of the propagator. We analyze the dynamics, in the limit  $N\rightarrow\infty$, of observables of a fixed number $n$ of particles, of extensive particle observables and their fluctuations, as well as of reservoir observables. We illustrate our results on the infinite mode Dicke  model and on various energy conserving models.
\end{abstract}


\section{Introduction and main results}
\label{Sec2}

We consider a system of $N$ (possibly distinct) quantum `particles' interacting with a `reservoir' quantum system. The Hilbert space associated with particle $j$ is ${\cal H}_j$ and ${\cal H}_\r$ is that of the reservoir. The Hamiltonian acts on the total Hilbert space
\begin{equation}
\label{int1}
{\cal H}^{(N)} = {\cal H}_1\otimes\cdots\otimes {\cal H}_N\otimes{\cal H}_\r
\end{equation}
and is given by
\begin{equation}
\label{mo1}
H_N = \sum_{j=1}^N h_j +H_\r +\frac{\lambda}{\sqrt N} \sum_{j=1}^N G_j\otimes B_j.
\end{equation}
Here, $h_j$ is the Hamiltonian of the $j$th particle (a short way of writing $\one_1\otimes\cdots\one_{j-1}\otimes h_j\otimes\one_{j+1}\cdots\otimes\one_N\otimes\bbbone_\r$ acting nontrivially on ${\cal H}_j$) and $H_\r$ is the reservoir Hamiltonian acting on ${\cal H}_\r$. The interaction is characterized by self-adjoint operators $G_j\in{\mathcal M}_j$, where 
\begin{equation}
{\cal M}_j={\cal B}({\cal H}_j),\qquad j=1,\ldots,N
\end{equation}
is the algebra of bounded operators on ${\cal H}_j$. The assumption that $G_j$ is bounded is not necessary for our approach but simplifies the exposition and is relevant in applications. We would like to include reservoirs consisting of free Bose particles and thus we do not want to restrict our focus on bounded interaction operators $B_j$. Rather, we assume only that $B_j\in {\cal M}_\r$ is a self-adjoint (possibly unbounded) operator on ${\cal H}_\r$, where
\begin{equation}
{\cal M}_\r={\cal L}({\cal H}_\r)
\end{equation}
is the set of linear operators on ${\cal H}_\r$. Of course, it is assumed that $H_N$ is self-adjoint.  For $1\le n\le N$, we set 
\begin{equation}
{\mathcal M}_{\le n} = {\mathcal M}_1\otimes\cdots\otimes {\mathcal M}_n.
\end{equation}
The Heisenberg dynamics is defined by
\begin{equation}
\label{mo72}
\tau^t_{\lambda,N} (A) = \e^{\i tH_N} A \e^{-\i tH_N},\qquad A\in{\cal M}_{\le N}\otimes {\cal B}({\cal H}_\r)
\end{equation}
and we denote the free dynamics ($\lambda=0$) by
\begin{equation}
\label{2.7}
A(t)\equiv \tau_{0,n}^t (A),\qquad  A \in {\mathcal M}_{\le n}\otimes {\cal B}({\cal H}_\r).
\end{equation}
The initial state is taken of the form
\begin{equation}
\label{mo2}
\omega_N = \mu_1\otimes\mu_2\otimes\cdots\otimes \mu_N \otimes\mu_\r,
\end{equation}
where $\mu_j$ is a state on ${\mathcal M}_j$ and $\mu_\r$ is a state on ${\cal B}({\cal H}_\r)$. We view a state, say $\omega$, as a normalized linear functional on observables (as is usual in the `algebraic' formulation of quantum theory). Equivalently, one might think of a state as a density matrix, say $\rho$. The two notions are linked by $\omega(A) = {\rm Tr} (\rho A)$. We are often interested in the dynamics of unbounded reservoir observables (such as the number of excitations in a Bose field) and so we extend the definitions \eqref{mo72} and \eqref{2.7} to $A\in{\cal M}_{\le N}\otimes {\cal M}_\r$. Generally,  $\tau^t_{\lambda,N}(A)$ is then an unbounded operator on ${\cal H}^{(N)}$ and we assume throughout that $\omega_N$ is sufficiently `regular' so that  $\omega_N(\tau^t_{\lambda,N}(A))$ is well defined, for all $N$ and all $t$. 

\bigskip
\noindent
{\bf Definitions} 
\begin{itemize}
	\item[{\bf (1)}]  
We call the system {\bf symmetric} if ${\cal H}_j = {\cal H}_\s$, $h_j=h$, $G_j=G$, $B_j=B$ and $\mu_j=\mu_\s$ for all $j=1,\ldots,N$, where $h$ and $G$ are a single particle Hamiltonian and an interaction operator, $B$ is a reservoir interaction operator and $\mu_\s$ is a state on ${\cal M}_\s= {\cal B}({\cal H}_\s)$. 

\item[{\bf (2)}] We call the system {\bf energy conserving} if $G_j(t)=G_j$ for all $t$ and all $j$.
\end{itemize}

\bigskip

Our goal is to find an expansion of the dynamics $\omega_N(\tau_{\lambda,N}^t(A))$ in powers of $N^{-1/2}$. To do so, we proceed as follows.

\begin{itemize}
\item[$\bullet$] We define coefficients $X_{\nu,N}, Y_{\nu,N}$ accompanying $N^{-\nu}$ and $N^{-\nu-1/2}$ in such an expansion. These are functionals  on observables which {\em still depend on $N$} and are analytic in $\lambda$ at $\lambda=0$. Their Taylor expansions are given in \eqref{c4} and \eqref{c5}. 

\item[$\bullet$] We introduce two conditions (A0) and (A1) and show in Theorem \ref{XYthm} that the functionals $X_{\nu,N}, Y_{\nu,N}$ are well defined (expressed by convergent Taylor series in $\lambda$) and that they are {\em uniformly bounded} in $N$.

\item[$\bullet$] We give in Theorem \ref{mainthm} the expansion of $\omega_N(\tau_{\lambda,N}^t(A))$ in terms of the $X_{\nu,N}$ and $Y_{\nu,N}$.

\item[$\bullet$] We show in Theorem \ref{mainthmsymmetric} that for symmetric systems,  $X_{\nu,N}$ and $Y_{\nu,N}$ have limits as $N\rightarrow\infty$, which are again analytic in $\lambda$ at $\lambda=0$ ({\em c.f.}  \eqref{mo36}, \eqref{mo60}). 
\end{itemize}

The coefficients $X_{\nu,N}$, $Y_{\nu,N}$ are constructed from the Dyson series expansion of the dynamics (in which the unperturbed part is generated by $H_N$ with $\lambda=0$). They are given by integrals over multi-commutators (see Section \ref{mechsect} for the mechanism). We define them now.  Let $A(t)\in{\cal M}_{\le n}\otimes{\cal M}_\r$, $r\ge 1$ and $t\ge t_1\ge t_2\ge\cdots\ge t_r$ be fixed. Let $(p_1,\ldots,p_N)$ be an $N$-tuple of integers $p_j\in\{0,1,\ldots\}$. We associate to $(p_1,\ldots,p_N)$ a set of operators, denoted ${\cal C}_r(p_1,\ldots,p_N)$, consisting of the collection of all $r$-fold multi-commutators
\begin{equation}
\label{mo78}
 T_t = \big[G_{j_r} (t_r) \otimes B_{j_r}(t_r),[\ldots,[G_{j_1} (t_1) \otimes B_{j_1}(t_1), A(t)]\ldots ]\big],
\end{equation}
in which each index $j_1,\ldots,j_r$ varies over the values $\{1,\ldots,N\}$, under the {\em constraint} that $p_1$ among the indices equal $1$, $p_2$ among them equal $2$, and so on, and  $p_N$ among them equal $N$. For each integer $\nu\ge 0$, set
\begin{eqnarray}
X_{\nu,N}(A(t)) &=& N^{\nu}\sum_{r\ge 2\nu,\, r\neq 0 \atop \rm even} (i\lambda)^r N^{-r/2}\sum_{p_1+\cdots+p_n=2\nu}
\ \  \sum_{p_{n+1}+\cdots +p_N=r-2\nu \atop \rm even}\nonumber\\
&&\qquad \times \int_0^t dt_1\cdots\int_0^{t_{r-1}}d t_r \ \sum_{T_t\in{\cal C}_r(p_1,\ldots,p_N)} \omega_N\big( T_t   \big).
\label{c4}
\end{eqnarray}
The first three sums are over integers $r, p_1,\ldots,p_N\in\{0,1,2,\ldots\}$ and the indication `even' means that the respective sums are taken only over even summation indices (even $r$ in the first one, even $p_{n+1},\ldots,p_{N}$ in the third one). We also define, for $\nu\ge 0$,
\begin{eqnarray}
Y_{\nu,N}(A(t)) &=& N^{\nu+1/2} \sum_{r\ge 2\nu+1 \atop \rm odd} (i\lambda)^r N^{-r/2}\sum_{p_1+\cdots+p_n=2\nu+1}
\ \  \sum_{p_{n+1}+\cdots+p_N=r-2\nu-1 \atop \rm even}\ \   \nonumber\\
&& \qquad \qquad \times\int_0^t dt_1\cdots\int_0^{t_{r-1}}d t_r\ \sum_{T_t\in{\cal C}_r(p_1,\ldots,p_N)}  \omega_N\big( T_t   \big).
\label{c5}
\end{eqnarray}
We point out that the terms $\sum_{p_1+\cdots +p_n=2\nu}\cdots \omega_N(T_t)$ in  \eqref{c4}, \eqref{c5} still depend on $N$, but not on $\lambda$. It is then apparent that \eqref{c4} and \eqref{c5} define functions of $\lambda$ analytic at $\lambda=0$. $X_{0,N}(A(t))$ has a zero of order two at $\lambda=0$ and for $\nu \ge 1$, $X_{\nu,N}(A(t))$ has a zero of order $2\nu$ at $\lambda=0$. $Y_{\nu,N}(A(t))$ has a zero of order $2\nu+1$ at $\lambda=0$, for $\nu\ge 0$.

\smallskip

The conditions (A0) and (A1) below serve to control the convergence of the series \eqref{c4} and  \eqref{c5}.  We present models satisfying them in Section \ref{satsect}. 

\begin{itemize}
\item[\bf (A0)] {\em {\bf Vanishing odd moment condition.} We assume that for every $j=1,\ldots,N$, every integer $k\ge 0$ and for all times $t_1,\ldots, t_{2k+1} \in [0,t]$, 
\begin{equation}
\label{mo4}
\mu_j \big( G_j(t_1)\cdots G_j(t_{2k+1})\big) =0.
\end{equation}
}
\end{itemize}
Given $A_\r\in{\cal M}_\r$, $t\ge 0$ and an integer $r\ge 1$, we define
\begin{eqnarray}
\beta_r(A_\r,t) &\equiv& \sup_{t_1,\ldots, t_r\in[0,t]} \ \max_{\sigma \in {\frak S}_r} \, \max_{1\le j\le r+1}\big| \mu_\r\big( B_{\sigma(1)}(t_1) \cdots  [\, {}_j A_\r(t)] \cdots B_{\sigma(r)}(t_r) \big) \big| \ \  \label{beta}\\ 
b(A_\r,t)&\equiv& \limsup_{r\rightarrow\infty} \frac{[\beta_r(A_\r,t)]^{1/r}}{\sqrt r},
 \label{A1}
\end{eqnarray}
where ${\frak S}_r$ is the group of permutations and the symbol $[\,{}_j A_\r(t)]$ means that $A_\r(t)$ is the $j$th factor inside the product. We make the following assumption. 
\begin{itemize}
\item[{\bf (A1)}] {\em Let $g = \sup_{j\ge 1} \| G_j\|$. We have
\begin{equation}
\label{c8}
2\sqrt{2\e}\, |\lambda| g t \,b(A_\r,t)  <1.
\end{equation}
}
\end{itemize}
The bound \eqref{c8} gives the time scale for which our results hold. We show in Section \ref{satsect} that for some reservoirs the bound is satisfied for all $\lambda$, $t\in\mathbb R$ (e.g. when $B$ is a bounded operator). For others the bound imposes the constraint $|\lambda| t\le C$ for some finite constant $C>0$ (e.g. when $B$ is a Bose field operator and $\mu_\r$ is quasi-free).

The following result shows that the functionals $X_{\nu,N}$ and $Y_{\nu,N}$ are well defined.
\begin{thm}
\label{XYthm}
The series \eqref{c4} and \eqref{c5} converge for any $n$ ($<N$) and any $A=A_\s\otimes A_\r \in {\mathcal M}_{\le n}\otimes {\mathcal M}_\r$. Moreover, 
\begin{itemize}
\item[(A)] The maps $\lambda\mapsto X_{\nu,N} (A(t))$ and $\lambda\mapsto Y_{\nu,N} (A(t))$ are analytic in a disc centered at $\lambda=0$, of radius $R$ having the $N$- and $\nu$-independent lower bound
\begin{equation}
\label{roc}
R \ge \frac{1}{2 \sqrt{2\e} \,g t \,b(A_\r,t)}. 
\end{equation}

\item[(B)] For $\nu\ge0$, we have 
 \begin{eqnarray}
 \label{Xbound}
 |X_{\nu,N}(A(t))| &\le&\|A_\s\| \frac{(2n|\lambda| g \,t)^{2\nu}}{(2\nu)!} S_\nu(A_\r,t),\\
 |Y_{\nu,N}(A(t))|& \le& \|A_\s\| \frac{(2n|\lambda| g \,t)^{2\nu+1}}{(2\nu +1)!} S_{\nu+1/2}(A_\r,t),
 \label{Ybound}
 \end{eqnarray}
where
\begin{equation}
\label{S}
S_\nu(A_\r,t) = \sum_{s\ge \delta_{0,\nu} }\frac{(2|\lambda|g \,t)^{2s}}{s!} \beta_{2(\nu+s)}(A_\r,t).
\end{equation}
The summation in \eqref{S} starts at $s=0$ if $\nu>0$ and at $s=1$ for $\nu=0$. The series \eqref{S} converges. 
\end{itemize}
\end{thm}
We point out that the upper bounds in \eqref{Xbound} and \eqref{Ybound} are {\em independent of $N$}. Denoting for a moment the general term in the series \eqref{S} by $\varkappa_s$, we use \eqref{c8} to obtain  $\limsup_s |\varkappa_s|^{1/s}\le 2\,\e\, b^2(A_\r,t)\, (2|\lambda| g t)^2$. This shows that the series \eqref{S} converges due to \eqref{c8}. The linear functionals $X_{\nu,N}$ and $Y_{\nu,N}$ are used to express the dynamics as follows.
\begin{thm}
\label{mainthm}
Let $A=A_\s\otimes A_\r \in {\mathcal M}_{\le n}\otimes {\mathcal M}_\r$, where $n<N$, and suppose that 
\begin{equation}
\label{thmbnd}
( n \e |\lambda| g\, t)^2 \limsup_{\nu\rightarrow\infty} \frac{[S_\nu(A_\r,t)]^{1/\nu}}{\nu^2} <1.
\end{equation} 
Then we have the expansion  
\begin{equation}
\label{c7}
\omega_N \big(\tau_{\lambda,N}^t (A)\big) = \omega_n (A(t))  + \sum_{\nu=0}^{\infty} N^{-\nu} X_{\nu,N}(A(t))+ N^{-\nu-1/2} Y_{\nu,N}(A(t)),
\end{equation}
where the series converges absolutely and uniformly in $N\ge 1$, and uniformly in $t$ for $t$ varying in compact sets.
\end{thm}

We show in  Lemma \ref{cor2} that  $X_{0,N}\upharpoonright_{{\cal M}_{\le n}}=0$ for any $n$, while $X_{0,N}$ does not vanish on ${\cal M}_\r$ (see Section \ref{resdyn}). In the symmetric situation (see the definition after \eqref{mo2}) the functionals $X_{\nu,N}$ an $Y_{\nu,N}$ have limits as $N\rightarrow\infty$.



\begin{thm}[Symmetric case]
\label{mainthmsymmetric}
Suppose that the system is symmetric and that the conditions of Theorem \ref{mainthm} hold. Then the limits
\begin{equation}
\label{mo32}
\lim_{N\rightarrow\infty} X_{\nu,N}(A(t)) = X_\nu(A(t)),\qquad \lim_{N\rightarrow\infty} Y_{\nu,N}(A(t)) = Y_\nu(A(t))
\end{equation}
exist and are uniform in $t$ for $t$ varying in compact sets. The limiting $X_\nu$, $Y_\nu$ are functionals analytic in $\lambda$ at $\lambda=0$. The observable $A_\s$ does not have to be symmetric with respect to permutation of particles for this result to hold.
\end{thm}

The Taylor expansions of $X_\nu$ and $Y_\nu$ are given in  \eqref{mo36} and \eqref{mo60}. Combining \eqref{mo32} with Theorem \ref{mainthm} yields the expansion
\begin{equation}
\label{mo61}
\omega_N \big(\tau_{\lambda,N}^t (A)\big) = \omega_n (A(t)) + \sum_{\nu=0}^{\infty} N^{-\nu} \big\{ X_\nu(A(t)) + o_N\big\}+ N^{-\nu-1/2} \big\{ Y_\nu(A(t)) +o_N\big\}, 
\end{equation}
with $o_N\rightarrow 0$ uniformly in $t$ for $t$ in compacta.  We have $X_0\upharpoonright_{{\cal M}_{\le n}}=0$ for any $n$ (see Lemma \ref{cor2}).

\subsection{Consequences of Theorems \ref{mainthm} and \ref{mainthmsymmetric} }

In this section, we present results that follow from our main theorems above. The proofs of all the Lemmas given below are presented in Section \ref{lemproofs}.

\subsubsection{The $n$-particle dynamics}
\label{partdyn}
\begin{lem}
	\label{cor2}
	Let $n<N$. We have $X_{0,N}\upharpoonright_{{\cal M}_{\le n}} =0$, meaning that in the limit $N\rightarrow\infty$, the dynamics on ${\cal M}_{\le n}$ is just the non-interacting one, 
	$$
	\lim_{N\rightarrow\infty} \omega_N\big( \tau_{\lambda,N}^t(A)\big) = \omega_n(A(t)),\qquad \forall A\in{\cal M}_{\le n}. 
	$$
\end{lem}

{\bf Remarks. } {\bf (1)} The fact that $X_{0,N}\upharpoonright_{{\cal M}_{\le n}} =0$ follows directly from \eqref{c4}. Indeed, for $\nu=0$, all the $p_1=\cdots=p_n=0$. This means that in the commutator $T_t$ in \eqref{c4} all the operators $G_j(t_j)$ act on particles $j$ with $j\ge n+1$, so they commute with any $A\in{\cal M}_{\le n}$ (see \eqref{mo78}). Hence $T_t=0$. In the special case of the Dicke model (see Section \ref{dickesection}), the result of Lemma \ref{cor2} was obtained in \cite{HL1}.

{\bf (2)} Lemma \ref{cor2} shows that the influence of the reservoir on the dynamics of any $n$-particle observable is negligible, as $N\rightarrow\infty$. However, $X_{0,N}\upharpoonright_{{\cal M}_\r}\neq 0$ and the reservoir experiences a non-trivial dynamics due to the presence of the particles (c.f. Section \ref{resdyn}).

\subsubsection{The reservoir dynamics} 
\label{resdyn}

Consider the symmetric case. According to Theorem \ref{mainthmsymmetric} we have 
\begin{equation}
\label{mo65}
\omega_N \big(\tau_{\lambda,N}^t(A_\r)\big) = \mu_\r(A_\r(t))+ X_0(A_\r(t)) +o_N,
\end{equation}
with $o_N\rightarrow 0$ as $N\rightarrow\infty$. The reservoir dynamics is not the free one in the large $N$ limit. The influence of the `particles' is given by the term $X_0$. The particles may themselves be seen as a `mean field reservoir', acting on the `system' $\r$.

The interaction term $X_0$ is analytic in $\lambda$ and we know its Taylor series, {\em c.f.} \eqref{mo36}. In the energy conserving situation we can resum the Taylor series to obtain the following result.

\begin{lem}
\label{lemma3}
Suppose that the system is symmetric and the dynamics is energy conserving. Let $\mu_{\rm HO}$ be the vacuum state of a quantum harmonic oscillator with creation and annihilation operators $a, a^*$, satisfying $[a,a^*]=\bbbone$ and set $\varphi=\frac{1}{\sqrt 2}(a+a^*)$. We have
\begin{equation}
\label{mo37.6}
\lim_{N\rightarrow\infty} \omega_N \big(\tau_{\lambda,N}^t(A_\r)\big) = \mu_{\rm HO}\otimes\mu_\r\Big( \e^{\i t (H_\r+2\sqrt{\varkappa} \varphi \otimes B)}  \big(\bbbone_{\rm HO}\otimes A_\r\big) \e^{-\i t (H_\r+2\sqrt{\varkappa} \varphi \otimes B)}\Big),
\end{equation}
where $\varkappa = \lambda^2\mu_\s(G^2)$.
\end{lem}

The result shows that the net effect of the $N\rightarrow\infty$ single particles on the reservoir is described by a single, zero frequency ($H_{\rm HO}=0$) quantum harmonic oscillator interacting linearly with the reservoir. The only trace of the original single particle system, in the limit $N\rightarrow\infty$,  is the variance $\mu_\s(G^2)$, determining the interaction strength between the oscillator and the reservoir $\r$.

\subsubsection{The leading orders}

We examine  the terms with $\nu=0$ in \eqref{c7} in more detail. The following result holds for systems which do not have to be symmetric.

\begin{lem}
\label{leadingorderthm}
	Assume the conditions of Theorem \ref{mainthm} and set $\mu_{\le n}=\mu_1\otimes\mu_2\otimes\cdots\otimes \mu_n$. We have  
	\begin{eqnarray}
	\lefteqn{	\omega_N \big(\tau_{\lambda,N}^t (A)\big) = \mu_{\le n} \big(A_\s(t)\big) \, \mu_\r(A_\r(t))}\nonumber\\
	&& - 2\frac{\lambda^2}{N}\,\mu_{\le n} \big(A_\s(t)\big)\sum_{j=n+1}^N\int_0^t ds \int_0^{s} ds' \, {\rm Re} \left\{ \mu_j(G_j(s')G_j(s)) \, \mu_\r\big(B_j(s') [B_j(s), A_\r(t)]\big) \right\} \nonumber \\
	&& -2\frac{\lambda}{\sqrt{N}} \sum_{j=1}^n \int_0^t d s\,  {\rm Im}\left\{ \mu_{\le n}\big(G_j(s) A_\s(t)\big)\mu_\r\big(B_j(s) A_\r(t)\big)\right\}\nonumber\\
	&& +O(\lambda^4 + \lambda^3/\sqrt{N} + 1/N) .
		\label{mo27}
	\end{eqnarray}
The remainder is uniform in $t$ varying in compact subsets of $|t| < (2\sqrt{2\e}\, |\lambda| g \,b(A_\r,t) )^{-1}$ and it is uniform in $t\in\mathbb R$ if $b(A_\r,t)=0$. 
\end{lem}
The second term on the right side of  \eqref{mo27} has a (finite) limit as $N\rightarrow\infty$ if, for instance, the model is asymptotically constant, {\em i.e.}, if $\mu_j\rightarrow\mu_\s$, $G_j\rightarrow G$ and $B_j\rightarrow B$ as $j\rightarrow\infty$. Indeed, 
\begin{eqnarray}
\lefteqn{
	-2\frac{\lambda^2}{N}\sum_{j=n+1}^N\int_0^t ds \int_0^{s} ds' \, {\rm Re} \left\{ \mu_j(G_j(s')G_j(s)) \, \mu_\r\big(B_j(s') [B_j(s), A_\r(t)]\big) \right\} }\label{mo30}\\
&=& -2\lambda^2 \int_0^t ds \int_0^{s} ds' \, {\rm Re} \left\{ \mu_\s(G(s')G(s)) \, \mu_\r\big(B(s') [B(s), A_\r(t)]\big) \right\}
+o(1/N),\nonumber
\end{eqnarray}
where $o(1/N)$ is a quantity that converges to zero as $N\rightarrow\infty$.  One can generalize \eqref{mo30} to the case where the quantities $\mu_j, G_j, B_j$ only converge in the sense of ergodic averages.

\subsubsection{The dynamics of intensive and fluctuation observables }
\label{fluctsect}

We show here that the average of intensive system observables evolves according to the free dynamics, but the reservoir induces system fluctuations. Consider the symmetric case (see after \eqref{mo2}) and let $A\in{\cal M}_\s$ be a single particle observable. The associated {\em intensive observable} is 
$$
A_{(N)}\equiv N^{-1} \sum_{n=1}^N A_n,\qquad A_n\equiv \bbbone_\s\otimes\cdots A\cdots\otimes\bbbone_\s\in {\cal M}_{\le N},
$$
where $A$ acts on the $n$-th factor. According to Lemma \ref{cor2} we have
\begin{equation}
\label{mo69}
\lim_{N\rightarrow\infty} \omega_N\big(\tau^t_{\lambda,N}(A_{(N)})\big) = \mu_\s(A(t)),
\end{equation}
meaning the expectation of intensive observables evolves according to the free dynamics, as $N\rightarrow\infty$. One may interpret \eqref{mo69}, which holds for all states $\mu_\s$, as a manifestation of the law of large numbers (for each fixed $t$). Namely the the sample average of the `random variables' $\tau^t_{\lambda,N}(A_n)$, $n=1,\ldots,N$,  converges to the limit $A(t)$. One defines the {\em fluctuation observable} \cite{Benatti}
$$
F_N(A,t) =\sqrt{N} \Big( \tau^t_{\lambda,N}(A_{(N)}) -\mu_\s(A(t))\Big).
$$
 The following result gives the limiting dynamics of fluctuation observables. 
 
\begin{lem}
\label{fluctlem}
Consider the symmetric situation. We have
\begin{eqnarray}
	\lefteqn{
\lim_{N\rightarrow\infty} \omega_N(F_N(A,t)) 
		 = \i\lambda \int_0^t \mu_\s([G(t_1), A(t)]) \mu_\r(B(t_1)) \, dt_1} \nonumber\\&&
	+\sum_{r\ge 3, {\rm \, odd}} \frac{(\i\lambda)^r}{(\frac{r-1}{2})!} \int_0^{t}d t_1\cdots \int_{0}^{t_{r-1}} dt_r \  \mu_\s \big([G(t_1), A(t)]\big) 
	\label{mo75.4}
	\\
	&& \qquad \times  {\sum_{j_2,\ldots, j_r} }^{\!\! *} \   \omega_{(r-1)/2}\Big( 
	\Big[ G_{j_r}(t_r) \otimes B(t_r), \big[\cdots ,[G_{j_2}(t_2)\otimes B(t_2), B(t_1)] \big]\Big]\Big),
	\nonumber
\end{eqnarray}
where the starred sum is over all indices $j_2,\ldots,j_r$ with the constraint that two among them equal $1$, two among them equal $2$, and so on, and two among them equal $(r-1)/2$. 
\end{lem}

Lemma \ref{fluctlem} shows that the fluctuations vanish if  $\mu_\r(B(t_1)\cdots B(t_{2k+1}))=0$ for all $k\ge 1$. This is in particular the case for a free Bose reservoir in a gauge invariant state $\mu_\r$ (e.g. the equilibrium state, or the vacuum, or a state with a definite number of excitations), coupled to the system via a field operator $B=\varphi(f)$. 

We now discuss an example of a system with non vanishing fluctuations. Consider each system particle to be a spin 1/2, with $h_j=\frac12\omega_0 \sigma_z$, and the reservoir to be a harmonic oscillator with $H_\r=\omega_\r \, a^*a$ in a coherent state $\mu_\r = \langle\alpha|\, \cdot\, |\alpha\rangle$, $\alpha\in\mathbb C$. Let the interaction operator be $G_j=\sigma_x$ and $B=\varphi=\frac{1}{\sqrt 2}(a^*+a)$.\footnote{One verifies that for this model, $\beta(\bbbone_\r, t)$, given in \eqref{beta}, grows like $\sqrt{r}$ for large $r$ and therefore $b(\bbbone_\r,t)=C$ for a finite constant. This means our analysis is valid for finite time intervals determined by \eqref{c8}.}  (This is the single mode Dicke maser model with initial field in a coherent state, {\em c.f.} Section \ref{dickesection}.) Then  $a|\alpha\rangle = \alpha |\alpha\rangle$ and $\mu_\r(B(s)) = 2{\rm Re}(\e^{\i\omega_\r s}\alpha)$. The lowest order of the fluctuation is then explicitly
\begin{equation}
	\label{fluc1}
\lim_{N\rightarrow\infty} \omega_N(F_N(A,t)) 
		= -4\lambda \int_0^t {\rm Im}\big( \mu_\s\big([\e^{\i s\omega_0}\sigma_+ , A(t)]\big) \big)  {\rm Re}(\e^{\i s\omega_\r}\alpha)\, ds +O(\lambda^3).
\end{equation}
Consider the spins to be in equilibrium at temperature $T$ and assume that the initial field coherent state is given by $\alpha\in{\mathbb R}$, and that the off-diagonal of the observable $A$ is also real, $A_{\uparrow\downarrow} = \langle\,\uparrow | A | \downarrow\, \rangle \in\mathbb R$. Then we calculate \eqref{fluc1} further to be
\begin{equation}
	\label{fluc2}
	\lim_{N\rightarrow\infty} \omega_N(F_N(A,t)) 
	= \frac{-4\lambda \alpha A_{\uparrow\downarrow}}{1+\e^{-\omega_0/T}} \sin(\omega_0 t) \Big(
\frac{\sin(\omega_0+\omega_\r)t}{\omega_0+\omega_\r} +\frac{\sin(\omega_0-\omega_\r)t}{\omega_0-\omega_\r}
	\Big)	
	+O(\lambda^3).
\end{equation}
This shows that only the coherences (off-diagonals in the energy basis) of the observable $A$ contribute to the fluctuations and for diagonal $A$ they vanish. Moreover, the fluctuations are oscillating in time, their onset is quadratic in time and the magnitude of the fluctuations is a decreasing function of the temperature $T\ge 0$, but still has a nonzero value for $T\rightarrow\infty$.

\subsection{Satisfying conditions (A0), (A1) and \eqref{thmbnd}}
\label{satsect}

\subsubsection{Condition (A0)}

\begin{itemize}
	\item[{\bf (1)}] Let the particle $j$ be described by a $d_j$-level system with Hamiltonian 
	$$
	h_j ={\rm diag}(e_j^{(1)},\ldots,e_j^{(d_j)}).
	$$ 
	Take for $G_j$ an operator inducing single step excitations and de-excitations, written in the diagonal basis of $h$ as
	\begin{equation}
	\label{mo31}
	G_j = \begin{pmatrix}
	0 & \gamma_{j,1} & 0 & \\
	\overline{\gamma_{j,1}} & 0 & \gamma_{j,2} & \\
	0 & \overline{\gamma_{j,2}} & 0 & \ddots&0\\
	&  & \ddots & \ddots & \gamma_{j,d_j-1}\\
	& & 0& \overline{\gamma_{j,d_j-1}}& 0
	\end{pmatrix},
	\end{equation}
	where the upper and lower diagonal consists of possibly nonzero numbers and all other entries vanish. Denote by $V_{\ell_1,\ell_2,\ldots}$ the span of the vectors $\phi_j^{(\ell_1)}, \phi_j^{(\ell_2)},\ldots$, where $\phi_j^{(\ell)}$ is the $\ell$-th excited state, i.e., $h_j\phi_j^{(\ell)} = e_j^{(\ell)}\phi_j^{(\ell)}$. Appying successively $G_j(t_1)$, $G_j(t_2)$, $G_j(t_3),\ldots$ to the state $\phi_j^{(1)}$  gives vectors belonging to the subspaces
	$$
	V_1 \mapsto V_2 \mapsto V_{1,3}\mapsto V_{2,4} \mapsto \cdots
	$$
	and it is manifest that $G_j(t_1)\cdots G_j(t_{2k+1})\phi_j^{(1)}\perp \phi_j^{(1)}$. The same holds for $\phi_j^{(1)}$ replaced by $\phi_j^{(\ell)}$ for any $\ell$. Therefore, the relation \eqref{mo4} is satisfied for $\mu_j(\cdot )= \langle{\phi_j^{(\ell)}},\ \cdot\ \phi_j^{(\ell)}\rangle$, any $\ell$. 
	
	\item[{\bf (2)}] A special case of the above example is $d_j=2$ and $G_j=\sigma_x$, the Pauli (spin flip) matrix.
	
	\item[{\bf (3)}]  If \eqref{mo4} holds for states $\mu_j^{(1)}, \mu_j^{(2)},\ldots,\mu_j^{(m)}$, then it holds for any of their mixture, $p_1\mu_j^{(1)}+p_2 \mu_j^{(2)}+\ldots+p_m\mu_j^{(m)}$. In particular, in the examples (1), (2) above, condition \eqref{mo4} is satisfied for the equilibrium states given by the density matrix $\rho_j\propto\e^{-\beta h_j}$, and more generally for any density matrix which is a function $\rho_j=f(h_j)$.
\end{itemize}

\subsubsection{Condition (A1)}

We introduce two classes (E1) and (E2) of examples to which we will refer in the sequel repeatedly.

\medskip

\noindent
{\bf Example (E1)}

Each $B_j$ is a bounded operator, with $g_\r\equiv \sup_j \|B_j\| <\infty$. Then we have $\beta_r(t)\le \|A_\r\| (g_\r)^r$ and consequently, $b(\lambda,t)=0$. It follows that \eqref{c8} is satisfied for all $t,\lambda\in\mathbb R$. 

\bigskip

\noindent
{\bf Example (E2)}

The reservoir describes a bosonic quantum field and the $B_j$ are field operators, i.e., $B_j=\varphi(f_j)$, $f_j\in L^2({
	\mathbb R}^3, d^3x)$ with $\varphi(f) = \frac{1}{\sqrt{2}}[a(f) +a^*(f)]$ and the dynamics is given by $B_j(t) = \varphi(\e^{\i t h_\r}f)$, for some self-adjoint one-particle Hamiltonian $h_\r$.
The field state $\mu_\r$ is a Gaussian (gauge-invariant, quasifree) state with two-point function $\mu_\r(B(s)B(s'))$. We prove the following result in Section \ref{lemproofs}. 
\begin{lem}
	\label{lemma1}
	Consider the example (E1). 
	{\em \bf (1)}  Let $A_\r\in{\mathcal M}_\r$. Then 
	$$
	\beta_r(A_\r,t)\le \frac{\e}{\sqrt{\pi}} \|A_\r\| (C(t)/\e)^{r/2}  r^{r/2} \mbox{\qquad and\qquad } b(A_\r,t) \le \sqrt{C(t)/\e},
	$$
	where $C(t) \equiv \max_{1\le i,j\le N} \max_{0\le s\le s'\le t} \big|\mu_\r\big(\varphi(\e^{\i s h_\r}f_i) \varphi(\e^{\i s' h_\r}f_j)\big) \big|$.
	
	\qquad {\em \bf (2)} Let $A_\r$ be a possibly unbounded reservoir operator, denote by $\widehat N$ the number operator and suppose that the state $\mu_\r$ carries at most $n_0$ particles. (More precisely, $\mu_\r( P(\widehat N>n_0) B_\r)=\mu_\r(B_\r P(\widehat N>n_0))=0$ for any $B_\r\in{\mathcal M}_\r$, where $P(N>n_0)$ is the spectral projection of $N$.) Then the upper bound for $\beta_r$ given in point $(1)$ above applies with $\|A_\r\|$ replaced by $\|A_\r P(N\le n_0+r/2)\|$ and the upper bound for $b$ is the same as the one given in point (1).

	\qquad {\em \bf (3)} Suppose $A_\r= \varphi(g_1)\cdots \varphi(g_k)$ is product of $k$ field operators. Then
	\begin{equation*}
	\label{mo10}
	\beta_r(A_\r,t) \le \frac{\e}{\sqrt \pi} \big(\frac{ C(A_\r,t)}{\e}\big)^{(r+k)/2} (r+k)^{(r+k)/2} \quad \mbox{and}\quad b(A_\r,t)\le \sqrt{C(A_\r,t)/\e},
	\end{equation*}
	where $
	C(A_\r,t) = \max_{f,f'\in \{f_1,\ldots,f_N,g_,\ldots,g_k\}} \max_{0\le s, s'\le t} \big|\mu_\r\big(\varphi(\e^{\i sh_\r}f) \varphi(\e^{\i s'h_\r}f')\big) \big|$. 
\end{lem} 

A sufficient parameter constraint for \eqref{c8} to hold is $8\lambda^2g^2t^2C<1$, where $C$ is the appropriate (time-dependent) constant given in Lemma \ref{lemma1} (1), (2) or (3).

\subsubsection{Condition \eqref{thmbnd}}

In example (E1), we  have $S_\nu(A_\r,t)\le \|A_\r\| (g_\r)^{2\nu}\e^{(2|\lambda| g g_\r  t)^2}$ and $\limsup_\nu [S_\nu(A_\r,t)]^{1/\nu}/ \nu^2=0$. It follows that \eqref{thmbnd} holds for arbitrary $t, \lambda\in\mathbb R$.

\medskip

For example (E2), we have the following result (proven in Section \ref{lemproofs}).
\begin{lem}
	\label{lemma2} Suppose that in the situations (1)--(3) of Lemma \ref{lemma1}, we have $16\lambda^2 g^2t^2 C(t)<1$ or $16\lambda^2g^2t^2 C(A_\r,t)<1$. Then  $\limsup_\nu [S_\nu(A_\r,t)]^{1/\nu}$ grows at most linearly in $\nu$, so $\limsup_\nu [S_\nu(A_\r,t)]^{1/\nu}/\nu^2=0$. Hence \eqref{thmbnd} does not introduce any additional bounds on $t,\lambda$. 
\end{lem}

\subsection{Embedding in previous work}

The literature on mean field limits of closed systems of interacting particles is huge. The topic has been a very active research field in physics and mathematical physics for many decades. We find Spohn's paper \cite{Spohn} particularly useful to understand, with little cumbersome technicality, the essence of the phenomena emerging in the mean field limit, and how to show them using the `BBGKY' hierarchy method. An excellent, more detailed account and overview of the literature is presented in \cite{BenedikterBook}. The literature on mean field limits of open quantum systems is less rich, to our knowledge, with the exception of the Dicke  model in various variations, whose thermodynamics and dynamics has been studied in great detail by many authors, for instance in \cite{HL1, davies, Hioe, Fetal, Mori}. The methods allowing to treat the Dicke model use its symmetries in an essential way and are very specific to the model at hand. A more general approach is taken in \cite{AL, AM}, where nonlinear evolutions for density matrices are examined. However, a rigorous derivation of markovian nonlinear dynamics emerging from the mean field limit has not been derived (except in explicitly solvable models). The problem is that one should control the two limits of small coupling ($\lambda$ small) and high complexity ($N$ large)  simultaneously. Our present work is an attempt at this, in that we derive a controlled expansion of the dynamics in both parameters $\lambda$ and $N$. We are not yet able to derive the limiting ($N\rightarrow\infty$) evolution equation in a `closed' form, say as a Hartree equation. Instead, here we only derive an expansion of the limiting dynamics in $\lambda$, which is generally not `resummable', except for energy conserving systems, as exemplified in our Lemma \ref{lemma3} and in the previous work \cite{MB}. On the other hand, our approach is a very direct analysis of the Dyson series expansion of the propagator and does not rely on many of the models' specifics.

\section{Illustrations}

\subsection{The Dicke  model}
\label{dickesection}

The Dicke (maser) model describes the interaction of (idealized, two-level) atoms with the quantized electromagnetic field. Its Hamiltonian is given by \cite{HL1, Hioe}
\begin{equation}
\label{d1}
H'_N=\frac{\omega_0}{2} \sum_{j=1}^N\sigma^z_j +  \nu \,a^*a +\frac{\lambda}{\sqrt N} \sum_{j=1}^N \sigma^x_j\otimes (a^*+a),
\end{equation}
where each atom has a transition (Bohr) frequency $\omega_0$, and where $\sigma^{x,z}_j$ are  Pauli matrices of the $j$th atom. In \eqref{d1}, the radiation is described by a single bosonic mode of frequency $\nu$, with associated creation and annihilation operators $a^*$ and $a$, satisfying $[a,a^*]=\bbbone$. The factor $1/\sqrt{N}$ is actually a factor $V^{-1/2}$, where $V$ is the volume of the cavity containing the atoms. Considering $V/N$ fixed leads to the prefactor in the interaction in \eqref{d1}. Often the rotating wave approximation is considered, where $\sigma^x_j\otimes (a^*+a)$ is replaced by $\sigma_j^+\otimes a + \sigma_j^-\otimes a^*$,  and then one can exploit the conservation of the total number of particles. We do not use this approximation here. A multimode model is given by \cite{Fetal}
\begin{equation}
\label{d2}
H''_N=\frac{\omega_0}{2} \sum_{j=1}^N\sigma^z_j +  \sum_{j=1}^N\nu_j \,a^*a +\frac{\lambda}{N} \sum_{i, j=1}^N  \sigma^x_j\otimes (a_i^*+a_i).
\end{equation}
In the limit of continuous modes the $a_i$ are replaced by $a(k)$, where $k\in{\mathbb R}^3$ is a continuous (momentum) parameter, satisfying $[a(k), a^*(l)] = \delta(k-l)$ (Dirac delta), one obtains a Hamiltonian of the form 
\begin{eqnarray}
H_N &=& \frac{\omega_0}{2} \sum_{j=1}^N\sigma^z_j + H_{\r} +\frac{\lambda}{\sqrt{N}} \sum_{j=1}^N \sigma^x_j\otimes \varphi(g),\label{d5} \\
H_\r &=& \int_{{\mathbb R}^3} |k| a^*(k) a(k) \d^3k\label{Ha}\\
B\equiv \varphi(g) &=& \frac{1}{\sqrt 2} \big( a^*(g) +a(g) \big),
\label{Be}
\end{eqnarray}
acting on the Hilbert space ${\cal H}_N = \otimes_{j=1}^N {\mathbb C}^2 \otimes {\mathcal F}$, where ${\mathcal F}$ is the bosonic Fock space over the one-particle Hilbert space $L^2({\mathbb R}^3,\d^3k)$ (Fourier representation). The value $g(k)$ of the {\em form factor} $g\in L^2({\mathbb R}^3,\d^3k)$ governs the strength of the coupling of mode $k$ to the collection of atoms. One factor $1/\sqrt N$ in \eqref{d2} is absorbed into the continuous creation and annihilation operators. The Hamiltonian \eqref{d5} with $\lambda/\sqrt N$ replaced by $\lambda/N$ (and in the rotating wave approximation) was considered by Davies in \cite{davies}. Davies' model is thus a weak coupling limit of \eqref{d5}, as was also remarked in \cite{Fetal}. We analyze the average field excitation
 \begin{equation}
 \label{d8}
 {\cal N}(t) = \lim_{N\rightarrow\infty} \omega_N\big( \tau_{\lambda,N}^t(
 \widehat N_\r) \big) = \mu_\r(\widehat N_\r) + X_0(\widehat N_\r)
 \end{equation} 
where $\widehat N_\r = \int_{{\mathbb R}^3} a^*(k)a(k)\d^3k$ is the number operator (see also \eqref{mo65}). We take the initial state the form \eqref{mo2} in which the field is in the vacuum $\mu_\r = |\Omega\rangle\langle\Omega|$ (we can deal with excited states in just the same manner) and, for some $p\in[0,1]$, we take
\begin{equation}
\label{d9}
\mu_j = p|\!\uparrow\rangle\langle\uparrow\!| +(1-p) |\!\downarrow\rangle\langle\downarrow\!|.
\end{equation}
We thus have (see \eqref{mo36})
\begin{equation}
\label{d10}
{\mathcal N}(t) = X_0(\widehat N_\r) = \sum_{q\ge 1} \frac{(-\lambda^2)^q}{q!} 
\int_0^t dt_1\cdots\int_0^{t_{2q-1}}d t_{2q} \  \sum_{T_t\in{\mathcal D}_q} \omega_q\big( T_t   \big) 
\end{equation}
Here, ${\mathcal D}_q$ is the class of all $(r=2q)$-fold multicommutators of the form \eqref{mo78} for which exactly two among the indices $j_1,\ldots,j_{2q}$ take each one of the values $1,\ldots,q$. The lowest order in $\lambda$ can be calculated directly either from \eqref{d10} or using Lemma \ref{leadingorderthm}, \eqref{mo27} and a few easy calculations. We obtain
\begin{equation} 
{\mathcal N}(t) = \lambda^2 \int_{{\mathbb R}^3}  |g(k)|^2 \Big[ p\frac{1-\cos(\omega_0-\omega)t}{(\omega_0-\omega)^2} + (1-p) \frac{1-\cos(\omega_0+\omega)t}{(\omega_0+\omega)^2}
\Big] \d^3k +O(\lambda^4),
\end{equation}
with a remainder uniform in $t$ for $|t| < (2|\lambda|\, \|g\|)^{-1}$. (Use Lemma \ref{lemma1} (2), to see that $b(\widehat N,t)\le \|g\|/\sqrt{2\e}$ .) In the parameter regime considered, the average number of field excitations is oscillating in time. This has also been observed in \cite{HL1} for $\lambda$ smaller than a critical value $\lambda_{\rm c}$. Beyond this regime an exponential increase in time of ${\mathcal N}(t)$ is expected (superradiance). Uncovering this behaviour might be difficult in the present setup, as one should include all orders of $\lambda$ and take $\lambda$ not too small. Furthermore, $n$-body observables show nontrivial dynamics in the superradiant phase (see for instance \cite{HL1}, Theorem 4.2), and so in view of Lemma \ref{cor2}, one might have to relax condition (A0) to capture nontrivial effects in this regime (maybe even in the photon field).

The thermodynamic properties and dynamics of the Dicke maser model have been studied in great detail in the references mentioned above (and many others). The analysis is based tightly on the specifics of the model (in particular, certain reductions due to conservation laws). The goal here was to illustrate how our general approach applies to the Dicke maser model, reproducing some of the previous results.

\subsection{Energy conserving models}

{\bf Single particle dynamics.\ } The system dynamics for the energy conserving and symmetric situation has been solved explicitly \cite{MB} and without imposing the vanishing odd moment condition. In that paper, each particle is given by a $d$-level system, the resevoir is an infinitely extended thermal, free bosonic quantum field and the interaction operator $B$ is the field operator $\varphi(g)=\frac{1}{\sqrt 2}(a^*(g) +a(g))$. One of the motivations for this model is the analysis of coherence and entanglement of qubits subject to a collective noise. It is shown in \cite{MB} that for each fixed time $t$, the reduced density matrix of $n$ particles (obtained by tracing out all other particles as well as the reservoir) converges as $N\rightarrow\infty$ to the $n$-fold tensor product of single particle density matrices,  $\rho_\infty(t)\otimes\cdots\otimes\rho_\infty(t)$. The limiting single particle density matrix obeys the quadratic evolution equation 
\begin{equation}
\label{d2.1}
\i \frac{d}{dt}\rho_\infty(t) = [h,\rho_\infty(t)] +\lambda^2 F(t)\, {\rm Tr}_2 [G\otimes G,\rho_\infty(t)\otimes\rho_\infty(t)],
\end{equation}
where $h$ and $G$ are the single particle Hamiltonian and the interaction operator ({\em c.f.} \eqref{mo1}) and $F(t)$ is an explicit complex valued function. ${\rm Tr}_2$ denotes the partial trace over the second factor. We now explain how the findings of \cite{MB} relate to the ones gotten here.

Writing the partial trace in \eqref{d2.1} as \begin{equation}
\label{d3}
{\rm Tr}_2 [G\otimes G,\rho_\infty(t)\otimes\rho_\infty(t)] = [G,\rho_\infty(t)] \,{\rm Tr}(\rho_\infty(t)G),
\end{equation}
and using the fact that $h$ and $G$ commute, it follows easily that  $\frac{d}{dt} {\rm Tr}\big(\rho_\infty(t) G\big) = 0$ for all $t$. This conservation law, when used in \eqref{d3} and \eqref{d2.1}, shows that in fact, the evolution \eqref{d2.1} takes the form
\begin{equation}
\label{d4}
\i \frac{d}{dt}\rho_\infty(t) = [h_{\rm eff}(t),\rho_\infty(t)] \qquad \mbox{with}\qquad h_{\rm eff}(t) =h +\lambda^2 \langle G\rangle_0 F(t)  G,
\end{equation}
where $\langle G\rangle_0={\rm Tr}(\rho_\infty(0) G)$ is the average of $G$ in the initial single particle state. Note that not only is the effective hamiltonian in \eqref{d4} time-dependent, but it depends on the initial condition as well (this is the hidden non-linearity). In the setting considered in the present paper, due to the vanishing odd moment condition (A0), we have $\langle G\rangle_0=0$ and hence $h_{\rm eff}=h$. That is, the particles evolve according to the free dynamics, as predicted by Lemma \ref{cor2}.

\bigskip

\noindent
{\bf Reservoir dynamics.\ } We are not aware that the reservoir dynamics has been considered before in the literature. Recall from Lemma \ref{lemma3} that in the large $N$ limit, any energy conserving model is equivalent to a single harmonic oscillator interacting with the reservoir. The only quantity of the particles playing a role is $\mu_\s(G^2)$ and we do not have to specify the particles system further. Here we will solve explicitly the reservoir dynamics given in Lemma \ref{lemma3}, \eqref{mo37.6} for a reservoir of free Bosons,  where $H_\r$ and $B$ are given by \eqref{Ha}, \eqref{Be}.  Denote by $\d E(x)$, $x\in\mathbb R$, the projection valued spectral measure of $\varphi$, so that $\varphi = \int_{\mathbb R} x \, \d E(x)$. We have 
\begin{equation}
\label{n1}
\e^{\i t (H_\r+2\sqrt{\varkappa} \varphi \otimes B)}  \big(\bbbone_{\rm HO}\otimes A_\r\big) \e^{-\i t (H_\r+ 2\sqrt{\varkappa} \varphi  \otimes B)} = \int_{\mathbb R} \d E(x) \otimes \e^{\i t (H_\r+\varphi(\xi g))}   A_\r  \e^{-\i t (H_\r+\varphi(\xi g))},
\end{equation}
where $\xi = 2\sqrt{\varkappa} x$ and on the right side, $\varphi(\xi g) = \xi\varphi(g)$ is the free Bose field operator smeared out with the form factor $\xi g(k)$, $k\in{\mathbb R}^3$. For a general form factor $f\in L^2({\mathbb R}^3)$ let $W(f)=\e^{\i\varphi(f)}$ be the Weyl operator. Using the standard relations $W(f)H_\r W(-f) = H_\r-\varphi(\i\omega f)+\frac12 \|\sqrt{\omega}f\|^2_2$ (with $\omega=|k|$) and $W(f)\varphi(g)W(-f) = \varphi(g) -{\rm Im}\langle f, g\rangle$, we readily verify the following relation, choosing $f(k)=\frac{\xi g(k)}{\i\omega}$,
\begin{equation}
\label{n2}
W(f) \e^{\i t(H_\r +\varphi(\xi g))} W(-f) = \e^{\i t H_\r} \, \e^{\frac i2 t \xi^2 \|g/\sqrt\omega\|^2_2}.
\end{equation}
(This is the `polaron transformation'.) Combining \eqref{n1} and \eqref{n2} yields
\begin{eqnarray}
\lefteqn{
e^{\i t (H_\r+2\sqrt{\varkappa} \varphi \otimes B)}  \big(\bbbone_{\rm HO}\otimes A_\r\big) \e^{-\i t (H_\r+ 2\sqrt{\varkappa} \varphi  \otimes B)} }\nonumber\\
&=& \int_{\mathbb R} \d E(x) \otimes W(-f) \e^{\i t H_\r}  W(f) A_\r W(-f)  \e^{-\i t H_\r} W(f)\nonumber\\
&=&\int_{\mathbb R} \d E(x) \otimes W\big((\e^{\i\omega t}-1)f\big) A_\r(t) W\big(-(\e^{\i\omega t}-1)f\big).
\label{n3}
\end{eqnarray}
In the last step we have used $\e^{\i tH_\r}W(f)\e^{-\i tH_\r} = W(\e^{\i t\omega}f)$ and the Weyl canonical commutation relations $W(f)W(h)=\e^{-\frac{\i}{2}{\rm Im}\langle f,h\rangle} W(f+h)$. Taking for $A_\r=W(h)$ a general Weyl operator ($h\in L^2({\mathbb R}^3)$), the second tensor factor on the right side of \eqref{n3} is $\e^{-\i {\rm Im} \langle f, (1-\e^{\i\omega t})h\rangle} W(\e^{\i\omega t}h) =\e^{-2\i\sqrt{\varkappa} x {\rm Re} \langle g, \frac{1-\e^{\i\omega t}}{\omega}h\rangle}W(\e^{\i\omega t}h)$. Using this information in \eqref{n3}, we find
\begin{equation}
\e^{\i t (H_\r+2\sqrt{\varkappa} \varphi \otimes B)}  \big(\bbbone_{\rm HO}\otimes W(h)\big) \e^{-\i t (H_\r+ 2\sqrt{\varkappa} \varphi  \otimes B)} = \e^{-2\i\sqrt{\varkappa} \, {\rm Re} \langle g, \frac{1-\e^{\i\omega t}}{\omega}h\rangle\varphi }\otimes W(\e^{\i\omega t}h).
\label{n4}
\end{equation}
The average in the harmonic oscillator vacuum state is
$$
\mu_{\rm HO}\big(  \e^{-2\i\sqrt{\varkappa} \, {\rm Re} \langle g, \frac{1-\e^{\i\omega t}}{\omega}h\rangle\varphi }   \big) = \e^{-\varkappa  ({\rm Re} \langle g, \frac{1-\e^{\i\omega t}}{\omega}h\rangle )^2}
$$
and so  combining \eqref{n4} with \eqref{mo37.6} gives the explicit limiting dynamics for the reservoir, 
\begin{equation}
\label{n5}
\lim_{N\rightarrow\infty} \omega_N(\tau_{\lambda,N}^t(W(h))) =  \e^{-\lambda^2 \mu_\s(G^2)  ({\rm Re} \langle g, \frac{1-\e^{\i\omega t}}{\omega}h\rangle )^2} \mu_\r(W(\e^{\i\omega t} h)).
\end{equation}
One can readily carry out this analysis for $A_\r={\widehat N}_\r$, the number operator of the reservoir Bose field. Let ${\mathcal N}(t) = \lim_{N\rightarrow\infty}\omega_N(\tau^t_{\lambda,N}(\widehat N_\r))$ be the average number of particles, as in Section \ref{dickesection}. Then one obtains
\begin{equation}
{\mathcal N}(t) = {\mathcal N}(0) +\lambda^2 \mu_\s(G^2)\, \| (\e^{\i\omega t}-1)g/\omega\|^2_2.
\end{equation}
This is an exact formula, in that there are no higher than quadratic order terms in $\lambda$ in the quantity ${\mathcal N}(t)$.  The number of particles is again oscillatory in time, for all values of $\lambda$. This indicates that for an energy conserving Dicke maser model, there is no superradiant phase transition. This is in contrast to the true (energy exchanging) Dicke model (see Section \ref{dickesection}), for which ${\mathcal N}(t)$ is oscillatory in time for $\lambda$ smaller than a critical value $\lambda_{\rm c}$, while for $\lambda>\lambda_{\rm c}$, ${\mathcal N}(t)$ increases exponentially in time \cite{HL1}.

\section{Proofs}

\subsection{The mechanism}
\label{mechsect}

For $A\in{\mathcal M}_{\le n}\otimes{\mathcal M}_\r$ we set $A(t)= \tau_{0,n}^t (A)$ and we expand in a Dyson series,
\begin{equation}
\label{eq:07}
\omega_N \big(\tau_{\lambda,N}^t (A)\big) = \omega_n \big(A(t)\big)  + \sum_{r=1}^{\infty} (i\lambda)^r \int_0^t dt_1 \cdots \int_0^{t_{r-1}} dt_r \, \omega_N \big( P_{r,N} (A(t))\big), 
\end{equation}
where, for $r\ge 1$,
\begin{equation}\label{eq:08}
P_{r,N}(A(t)) = N^{-r/2} \sum_{j_1,  j_2, \cdots,  j_r =1}^N \big[G_{j_r} (t_r) \otimes B_{j_r}(t_r),[\ldots,[G_{j_1} (t_1) \otimes B_{j_1}(t_1), A(t)]\ldots ]\big].
\end{equation}
It is convenient to make a change of variables in the sum \eqref{eq:08}. Given a fixed $r$-tuple $(j_1,\ldots,j_r)\in\{1,\ldots,N\}^r$ and a number $1\le k\le N$, define $0\le p_k\le r$ to be the number of $j$s in the tuple which equal to $k$. We have $p_1+\cdots +p_N=r$. To a given $(j_1,\ldots,j_r)$ there corresponds a unique $(p_1,\ldots,p_N)\in\{0,\ldots,r\}^N$, while a given $(p_1,\ldots,p_N)$ is associated to exactly 
\begin{equation}
\binom{r}{p_1,\ldots,p_N} \equiv \frac{r!}{(p_1)!\cdots (p_N)!}
\label{mm04}
\end{equation}
distinct  $(j_1,\ldots,j_r)\in\{1,\ldots,N\}^r$. We denote by ${\cal C}_r(p_1,\ldots,p_N)$ the set of all $\binom{r}{p_1,\ldots,p_N}$ summands in \eqref{eq:08} with different values of $(j_1,\ldots,j_r)$ associated to the same $(p_1,\ldots,p_N)$. Using this change of variables in \eqref{eq:08} results in the expression
\begin{eqnarray}\label{eq:010}
P_{r,N}(A(t)) &=& N^{-r/2} \, \sum_{p_1 + \cdots + p_N = r} \, \, \sum_ {T_t \in \mathcal{C}(p_1,\ldots, p_N) } T_t,
\end{eqnarray}
which defines the terms $T_t$. We split the sum over the $p_1,\ldots,p_N$ to obtain
\begin{eqnarray}\label{eq:011}
P_{r,N}(A(t)) &=& N^{-r/2} \,  \sum_{p=0}^{r} \, \, \sum_{p_1 + \cdots + p_n = r-p} \, \,  \sum_{p_{n+1} + \cdots + p_N = p} \, \, \sum_ {T_t \in  \mathcal{C}(p_1,\ldots,p_N) } T_t. 
\end{eqnarray}
Here, $p=p_{n+1}+\cdots +p_N\in\{0,\ldots r\}$ is the sum of the  powers of all factors $G$ in $T_t$ acting on particles with index larger than $n$. By the vanishing odd moment condition, all terms with odd values of any of the  $p_{n+1},\ldots, p_N$ give a vanishing contribution to \eqref{eq:011}. Set $p_j=2q_j$ for $j=n+1,\ldots,N$. Then
\begin{eqnarray}\label{eq:011.1}
P_{r,N}(A(t)) &=& N^{-r/2} \,  \sum_{p=0}^{r}{}' \, \, \sum_{p_1 + \cdots + p_n = r-p} \, \,  \sum_{q_{n+1} + \cdots + q_N = p/2} \, \, \sum_ {T_t \in  \mathcal{C}(p_1,\ldots,p_n,2q_{n+1},\ldots, 2q_N) } T_t\qquad  
\end{eqnarray}
where the prime ${}'$ indicates that we only sum over even numbers.  

To get an idea of the $N$-dependence of \eqref{eq:011.1}, we estimate the number of terms,
\begin{equation}
\label{c1}
\sum_ {T_t \in  \mathcal{C}(p_1,\ldots, 2q_N) } 1 \ \ =\ \   \binom{r}{p_1,\ldots,2q_N} <  \frac{r!}{(p_1)!\cdots (p_n)!} \frac{1}{(p/2)!} \binom{p/2}{q_{n+1},\ldots,q_N}.
\end{equation}
Then, since 
\begin{equation}
\label{c2}
\sum_{q_{n+1}+\cdots +q_N=p/2} \binom{p/2}{q_{n+1},\ldots,q_N} = (N-n)^{p/2} \sim N^{p/2},
\end{equation}
we expect an upper bound
\begin{equation}
\label{c3}
|\omega\big(P_{r,N}(A(t))\big)| \ \lesssim\  \sum_{p=0}^{r}{}' \ \  N^{(p-r)/2} \!\!\sum_{p_1 + \cdots + p_n = r-p} f(p_1,\ldots,p_n,r,p,t),
\end{equation}
where $f$ does not depend on $N$. For $r$ even, the terms generated in \eqref{c3} are of the orders $N^{-\frac r2}, N^{-\frac r2+1},\ldots,N^0$, all powers being {\em integers}. For $r$ odd, the terms generated in \eqref{c3} are of the orders $N^{-\frac r2}, N^{-\frac r2+1},\ldots,N^{-\frac12}$, all powers being {\em half integers}. Having in mind an expansion of \eqref{eq:07} in negative powers of $N$, we can ask which terms will give a contribution to $N^{-\nu}$. For an integer $\nu\ge 0$, such a term will be associated with even $r$ only. From \eqref{c3}, $(p-r)/2=-\nu$, or, $p=r-2\nu$. Similarly, terms with $N^{-\nu -1/2}$, $\nu\ge 0$, are associated with odd $r$ in \eqref{c3}, such that $(p-r)/2=-\nu-1/2$, i.e., $p=r-2\nu-1$. It follows that the quantities $X_{\nu,N}$ and $Y_{\nu,N}$, defined in \eqref{c4} and \eqref{c5} give the contributions to the series on the right side of \eqref{eq:07} of order $N^{-\nu}$ and $N^{-\nu-1/2}$, respectively.

\subsection{Proof of Theorem \ref{XYthm}}

We first prove (B).  The bound $|\omega_N(T_t)|\le \|A_\s\| (2g)^r \beta_r(A_\r, t)$ follows directly from the form of $T_t$ as an $r$-fold multi-commutator ({\em c.f.} \eqref{eq:08}, \eqref{eq:010}). Using the equality in \eqref{c1} gives
\begin{equation}
\label{p1}
\Big| \int_0^t dt_1\cdots\int_0^{t_{r-1}}d t_r \ \sum_{T_t\in{\cal C}_r(p_1,\ldots,p_N)}\omega_N\big( T_t   \big)\Big| \le \|A_\s\| \frac{(2gt)^r}{r!} \beta_r(A_\r,t)\, \binom{r}{p_1,\ldots,p_N}.
\end{equation}
Setting $p_j=2q_j$, for $n+1\le j\le N$, we get
\begin{equation}
\label{p2}
\binom{r}{p_1,\ldots,p_N} = \frac{r!}{(p_1)!\cdots (p_n)!}\frac{1}{(r/2-\nu)!}  \binom{r/2-\nu}{2q_{n+1},\ldots, 2q_N}
\end{equation}
and the latter multinomial is bounded above by $\binom{r/2-\nu}{q_{n+1},\ldots, q_N}$. We have
\begin{equation}
\label{p3}
\sum_{q_{n+1}+\cdots +q_N=r/2-\nu}  \binom{r/2-\nu}{q_{n+1},\ldots, q_N}
= (N-n)^{r/2-\nu}.
\end{equation}
Using \eqref{p1}-\eqref{p3} in \eqref{c4} yields
\begin{equation}
\label{p4}
|X_{\nu,N}(A(t))| \le \|A_\s\|  \sum_{r\ge 2\nu {\rm \, even},\, r\neq 0}\frac{(2|\lambda|g t)^r}{(r/2-\nu)! (2\nu)!}\, \beta_r(A_\r,t) \!\! \sum_{p_1+\cdots+p_n=2\nu} \binom{2\nu}{p_1,\ldots,p_n}.
\end{equation}
The last sum on the right side of \eqref{p4} equals $n^{2\nu}$ (c.f. \eqref{c2}). Finally, we make a change of variable $s=r/2-\nu$ in \eqref{p4} to arrive at the bound \eqref{Xbound}. The bound \eqref{Ybound} is obtained in the same way.

Now we prove (A). We write \eqref{c4} in the form $X_{\nu,N}(A(t)) = \sum_{r\ge 2\nu {\rm \, even},\, r\neq 0} \lambda^r\,  a_r$ with the obvious identification of the Taylor coefficients $a_r$. The radius of convergence is given by $R_\nu = 1/\limsup_r|a_r|^{1/r}$. Using the same bounds as in the proof of (A) above, we readily get
$|a_r| \le \|A_\s\| \frac{n^{2\nu}}{(2\nu)!}  \frac{(2gt)^r}{(r/2-\nu)!} \beta_r(A_\r,t)$, so that
$$
\limsup_r |a_r|^{1/r} \le 2 g t \limsup_r\big[\frac{\beta_r(A_\r,t)}{(r/2-\nu)!} \big]^{1/r} = 2 g t \limsup_r\big[\frac{\beta_r(A_\r,t)}{(r/2)!} \big]^{1/r}.
$$
Using Stirling's bound $(r/2)! \ge \sqrt{2\pi} (r/2)^{r/2+1/2} \e^{-r/2}$ then implies that
$$
\limsup_r |a_r|^{1/r} \le 2 \sqrt{2\e}  gt  \limsup_r \frac{[\beta_r(A_\r,t)]^{1/r}}{\sqrt{r}}.
$$
This shows \eqref{roc}. The bound on the radius of convergence for $Y_{\nu,N}(A(t))$ is obtained in the same way.  \hfill $\blacksquare$

\subsection{Proof of Theorem \ref{mainthm}}

The expression for the left side of \eqref{c7} is given by \eqref{eq:07} and \eqref{eq:011}. We split the sum \eqref{eq:07} into two, one sum for $r$ even and one for $r$ odd. As indicated before \eqref{c4}, for the even part, we make a change of variables, passing from $(r,p)$ to $(\nu,r)$, where $\nu=(r-p)/2$. For the sum with $r$ odd, we pass from $(r,p)$ to $(\nu,r)$, with $\nu=(r-p-1)/2$. This leads (formally) to the expansion \eqref{c7}. If the series $\sum_{\nu\ge 0} \{ |X_{\nu,N}(A(t))| +|Y_{\nu,N}(A(t))|\}$ converges uniformly in $N\ge 1$, then this rearrangement does not affect the value of the series and \eqref{c7} holds. We now show that $\sum_{\nu\ge 0} \{ |X_{\nu,N}(A(t))| +|Y_{\nu,N}(A(t))|\}$ converges uniformly in $N\ge 1$. Consider the bound \eqref{Xbound}. We have 
$$
\sum_{\nu\ge 0 } |X_{\nu,N}(A(t))| \le \|A_\s\| \sum_{\nu\ge 0}\frac{(2n|\lambda| g t)^{2\nu}}{(2\nu)!} S_\nu(A_\r,t),\\
$$ 
and the right side converges if $(2n|\lambda| \,g t)^2 \limsup_\nu [S_\nu(A_\r,t)/(2\nu)!]^{1/\nu}<1$. From Stirling's estimate, we have $(2\nu)! \ge \sqrt{2\pi} (2\nu)^{2\nu+1/2}\e^{-2\nu}$ and hence $\limsup_\nu [S_\nu(A_\r,t)/(2\nu)!]^{1/\nu} \le \frac{\e^2}{4}  \limsup_\nu S_\nu(A_\r,t)^{1/\nu}/\nu^2$. Similarly, one sees that $\sum_{\nu\ge 0 } |Y_{\nu,N}(A(t))|$ converges. This shows \eqref{c7}.  \hfill $\blacksquare$

\subsection{Proof of Theorem \ref{mainthmsymmetric}} For fixed $r$ and $\nu$, consider the sum over the $p_{n+1}+\cdots +p_N=r-2\nu$ in the general term of the series \eqref{c4} defining $X_{\nu,N}(A(t))$.  We split this sum as
\begin{equation}
\label{mo33}
\sum_{p_{n+1}+\cdots +p_N=r-2\nu \atop \rm even}  = \sum_{p_{n+1}+\cdots +p_N=r-2\nu\atop \rm even}^{\ \ \ \ \  *} + \sum_{p_{n+1}+\cdots +p_N=r-2\nu \atop \rm even}^{\ \ \ \ \ **},
\end{equation}
where in the first sum (*) on the right side, all $p_j\in\{0,2\}$ and in the second one (**), some $p_j\ge 4$ (recall that the $p_j$ are even). We now show that only the sum with the single star in \eqref{mo33} contributes to the expression \eqref{c4} in the limit $N\rightarrow\infty$. From the relations\footnote{The value of the first sum is the Taylor coefficient in front of $x^T$ of the function $(1+x+x^2+\cdots)^L$, i.e., $\frac{1}{T!}\frac{d^T}{dx^T}|_{x=0}(1-x)^{-L}$. The value of the second sum is the Taylor coefficient in front of $x^T$ of the function $(1+x)^L$, i.e., $\frac{1}{T!}\frac{d^T}{dx^T}|_{x=0}(1+x)^{L}$.}
\begin{equation}
\label{mo49}
\sum_{q_1+\cdots +q_L= T} 1 = {L+T-1 \choose T}\qquad \mbox{and} \qquad \sum_{q_1+\cdots +q_L=T}^{\ \ \ \ \ *} 1 = {L\choose T},
\end{equation}
where the $q_j=0,1,\ldots$ and the star in \eqref{mo49} means that we sum only over values $q_j\in\{0,1\}$, we deduce that ($L=N-n$, $T=r/2-\nu$)
\begin{eqnarray}
\lefteqn{
	\sum_{p_{n+1}+\cdots +p_N=r-2\nu \atop \rm even}^{\ \ \ \ \ **} 1 = {L+T-1\choose T} - {L\choose T}}\label{mo50} \\
&=&\frac{L^T}{T!}\big\{ (1+\tfrac{T-1}{L}) (1+\tfrac{T-2}{L}) \cdots (1+\tfrac{1}{L}) - (1-\tfrac{1}{L}) (1-\tfrac{2}{L}) \cdots (1-\tfrac{T-1}{L})
\big\}\nonumber\\
&=& N^{r/2-\nu} o_N,
\nonumber
\end{eqnarray}
where $o_N\rightarrow 0$ as $N\rightarrow\infty$. The general term in the series \eqref{c4} carries a factor $N^{\nu-r/2}$ and therefore for each fixed $r$, the part of the summand in \eqref{c4} associated with the doubly starred sum (c.f. \eqref{mo33}) converges to zero as $N\rightarrow\infty$.  It follows (from the Dominated Convergence Theorem for the series \eqref{c4}) that 
\begin{eqnarray}
\label{mo34}
\lim_{N\rightarrow\infty} X_{\nu,N}(A(t)) &=& \lim_{N\rightarrow\infty} \sum_{r\ge 2\nu ,\, r\neq 0 \atop \rm even} (i\lambda)^r N^{\nu-r/2}\sum_{p_1+\cdots+p_n=2\nu}
\ \  \sum_{p_{n+1}+\cdots +p_N=r-2\nu \atop \rm even}^{\ \ \ \ \ *} \nonumber\\
&&\qquad \times \int_0^t dt_1\cdots\int_0^{t_{r-1}}d t_r \ \sum_{T_t\in{\cal C}_r(p_1,\ldots,p_N)} \omega_N\big( T_t   \big).
\end{eqnarray}
Due to the invariance of $\omega_N(T_t)$ with respect to permutation of any of the particle indices $j=n+1,\ldots,N$, the value of $\omega_N\big( T_t   \big)$ is the {\em same} for every  one of the configurations  $(p_{n+1},\ldots,p_N)$ in the starred sum of \eqref{mo34}. (This does not hold for $j\le n$ since the observable $A_\s$ acts on the first $n$ particles.) We may thus set $p_{n+1}=\cdots=p_{n+r/2-\nu}=2$ and $p_j=0$ for $j=n+r/2-\nu+1,\ldots,N$, and take this term with the multiplicity ${N-n\choose r/2-\nu}$, which is the number of terms in the sum according to \eqref{mo49}. In other words,
\begin{equation}
\label{mo35}
\sum_{p_{n+1}+\cdots +p_N=r-2\nu \atop \rm even }^{\ \ \ \ \ *} \ \sum_{T_t\in{\cal C}_r(p_1,\ldots,p_N)} \omega_N\big( T_t   \big)= \textstyle {N-n \choose r/2-\nu}  \displaystyle \sum_{T_t\in{\cal D}_r(p_1,\ldots,p_n)} \omega_{n+r/2-\nu}\big( T_t   \big),
\end{equation}
where ${\cal D}_r(p_1,\ldots,p_n)$ is the set of all $r$-fold multicommutators \eqref{mo78}, where $p_j$ among the indices $j_1,\ldots, j_r$ equal $j$, for $j=1,\ldots,n$ (under the additional constraint that $p_1+\cdots+p_n=2\nu$), and two among the indices $j_1,\ldots, j_r$ equal each one of the values $n+1,\ldots,n+r/2-\nu$.
Combining \eqref{mo34} and \eqref{mo35} gives 
\begin{eqnarray}
\lefteqn{
	\lim_{N\rightarrow\infty} X_{\nu,N}(A(t))} \nonumber \\
&=&  \sum_{r\ge 2\nu,\, r\neq 0 \atop \rm even} \frac{(i\lambda)^r}{(r/2-\nu)!} \sum_{p_1+\cdots+p_n=2\nu}\ 
\int_0^t dt_1\cdots\int_0^{t_{r-1}}d t_r \  \sum_{T_t\in{\cal D}_r(p_1,\ldots,p_n)} \omega_{n+r/2-\nu}\big( T_t   \big) \nonumber \\
&\equiv& X_\nu(A(t)). \label{mo36}
\end{eqnarray}
The same argument applies to $Y_{\nu,N}(A(t))$, \eqref{c5}, and yields
\begin{eqnarray}
\lefteqn{
	\lim_{N\rightarrow\infty} Y_{\nu,N}(A(t)) = \sum_{r\ge 2\nu+1 \atop \rm odd}\frac{(i\lambda)^r}{(r/2-\nu-1/2)!} }\nonumber \\
&\times  &  \sum_{p_1+\cdots+p_n=2\nu+1}\ 
\int_0^t dt_1\cdots\int_0^{t_{r-1}}d t_r \  \sum_{T_t\in{\cal E}_r(p_1,\ldots,p_n)} \omega_{n+r/2-\nu-1/2}\big( T_t   \big)\nonumber \\
&\equiv& Y_\nu(A(t)). \label{mo60}
\end{eqnarray}
where ${\cal E}_r(p_1,\ldots,p_n)$ denotes the set of all multicommutators as in the sum of \eqref{eq:08} with the following constraint: $p_j$ among the indices $j_1,\ldots,j_r$ equal $j$, for $j=1,\ldots,n$ (with the constraint $p_1+\cdots +p_n=2\nu+1$) and two  among the indices $j_1,\ldots,j_r$ equal each of the values $n+1,\ldots, n+r/2-\nu-1/2$.

Using the bound \eqref{Xbound} one readily sees that $\sum_{\nu=0}^\infty N^{-\nu} [X_{\nu,N}(A(t)) - X_\nu(A(t))] \rightarrow 0$ in the limit $N\rightarrow\infty$. The analogous result holds for $Y_{\nu,N}$. \eqref{mo61} follows. \hfill $\blacksquare$

\subsection{Proofs of lemmas}
\label{lemproofs}

\subsubsection{Proof of Lemma \ref{lemma1}} According to Wick's theorem,
\begin{equation}
\label{c10}
\big| \mu_\r\big( B(t_{\sigma(1)}) \cdots B(t_{\sigma(r)}) \big) \big| \le \sum_{{\rm pairings }\, \sigma} \ \prod_{j=1}^{r/2}\  \big| \mu_\r\big(B(t_j) B(t_{\sigma(j)}) \big) \big|  \le  \frac{r!}{2^{r/2} ( r/2)!} C(t)^{r/2},
\end{equation}
where the sum is over $ \frac{r!}{2^{r/2}(r/2)!}$ pairings. For $r$ odd, $\mu_\r\big( B(t_{\sigma(1)}) \cdots B(t_{\sigma(r)}) \big)=0$. 

We now prove (1). We use the Cauchy-Schwarz inequality for states, $|\mu(XAY) | \le \|A\| \sqrt{\mu(XX^*) \mu(Y^*Y)}$, to estimate
\begin{eqnarray}
\lefteqn{
	\big| \mu_\r\big( B_{\sigma'(1)}(t_{\sigma(1)}) \cdots  [\, {}_j A_\r(t)] \cdots B_{\sigma'(r)}(t_{\sigma(r)}) \big) \big|} && \nonumber\\
&\le &  \|A_\r\|\  \big[ \mu_\r\big( B_{\sigma'(1)}(t_{\sigma(1)}) \cdots B_{\sigma'(j-1)}(t_{\sigma(j-1)})\,  B^*_{\sigma'(j-1)}(t_{\sigma(j-1)}) \cdots B^*_{\sigma'(1)}(t_{\sigma(1)}) \big) \big]^{1/2} \nonumber\\
&& \times \big[ \mu_\r\big( B^*_{\sigma'(r)}(t_{\sigma(r)}) \cdots B^*_{\sigma'(j)}(t_{\sigma(j)})\,  B_{\sigma'(j)}(t_{\sigma(j)}) \cdots B_{\sigma'(r)}(t_{\sigma(r)}) \big) \big]^{1/2}.
\label{mo64}
\end{eqnarray}
The right hand side of \eqref{mo64} is estimated using Wick's theorem, see \eqref{c10}, yielding 
\begin{eqnarray}
\lefteqn{
	\big| \mu_\r\big( B_{\sigma'(1)}(t_{\sigma(1)}) \cdots  [\, {}_j A_\r(t)] \cdots B_{\sigma'(r)}(t_{\sigma(r)}) \big) \big|} && \nonumber\\
&\le &  \|A_\r\|  (C(t)/2)^{r/2} \left\{ \frac{( 2(j-1) )!}{(j-1)!} \frac{( 2(r-j+1))!}{(r-j+1)!} \right\}^{1/2}.
\label{mo7}
\end{eqnarray}
With the usual Stirling approximations,
\begin{equation}
\label{stirling}
\sqrt{2\pi} \,n^{n+1/2} \e^{-n} \le n! \le \e\, n^{n+1/2}\e^{-n},
\end{equation}
valid for all integers $n\ge 1$, we obtain
\begin{equation}
\label{stirling2}
(2n)!/n! \le \frac{e}{\sqrt \pi} (4/e)^n n^n
\end{equation}
and hence
\begin{equation}
\label{mo8}
\frac{( 2(j-1) )!}{(j-1)!} \frac{( 2(r-j+1))!}{(r-j+1)!} \le \frac{\e^2}{\pi} (4/\e)^r \ell^\ell (r-\ell)^{r-\ell},
\end{equation}
where $\ell =j-1$. The function $\ell\mapsto \ell^\ell (r-\ell)^{r-\ell} = (\frac{\ell}{r-\ell})^\ell (r-\ell)^r$ is readily seen to be maximal at $\ell=r/2$ (use simple calculus), where this function takes the value $(r/2)^r$. Combining the bound $\ell^\ell (r-\ell)^{r-\ell}\le(r/2)^r$ with \eqref{mo8} and \eqref{mo7} yields the upper bound on $\beta_r(A_\r,t)$ given in (1) of Lemma \ref{lemma1}. The upper bound on $b(A_\r,t)$ is then immediate from the definition \eqref{c8}.

The proof of (2) is obtained in the same way as (1). Indeed, since $\mu_\r$ has at most $n_0$ particles, and each $B$ can produce at most one particle, we may use the bound \eqref{mo64} with $\|A\|$ replaced by $\|A\,  P(\widehat N\le n_0+r/2)\|$.

Next we prove statement (3). We simply have to estimate $\mu_\r$ applied to a product of $r+k$ field operators. Wick's theorem gives the bound \eqref{c10}, 
\begin{equation}
\label{mo9}
\beta_r(A_\r,t) \le \frac{1}{2^{(r+k)/2}} \,  \frac{(r+k)!}{((r+k)/2)!} C(A_\r,t)^{(r+k)/2},
\end{equation}
Using the bound \eqref{stirling2} with $n=r+k$ yields the upper bound on $\beta_r(A_\s,t)$ in (2) of the lemma. The upper bound on $b(A_\r,t)$ is then immediate from the definition \eqref{c8}.

 \hfill$\blacksquare$

\subsubsection{Proof of Lemma \ref{lemma2}} Since the bounds on $\beta_r(A_\s,t)$ in (1), (2) and (3) of Lemma \ref{lemma1} have the same form, it suffices to give the proof in the case (3), for $k$ even. We obtain (recall \eqref{S})
\begin{eqnarray}
S_\nu(A_\r,t) &\le&  \frac{\e}{\sqrt\pi} (C/\e)^{\nu+k/2} (2\nu+k)^{\nu +k/2}  \nonumber  \\
&&+\frac{\e}{\sqrt\pi} (C/\e)^{\nu+k/2} \sum_{s\ge 1} \frac{(C\varkappa^2/\e)^s}{s!} \big( 2(\nu+s)+k\big)^{\nu+s+k/2}\nonumber \\
&\le&  \frac{\e}{\sqrt\pi} (C/\e)^{\nu+k/2} (2\nu+k)^{\nu +k/2}  \nonumber  \\
&& +\frac{\e}{\sqrt{2}\, \pi} (C/\e)^{\nu+k/2} \sum_{s\ge 1} (C\varkappa^2)^s \big( 2 +(2\nu+k)/s\big)^s \big( 2(\nu+s)+k\big)^{\nu+k/2} \qquad
\label{mo20}
\end{eqnarray}
where $\varkappa = 2|\lambda| gt$, $C=C(A_\r,t)$ and in the second step, we used Stirling's bound \eqref{stirling} for $s!$, $s\ge 1$. Using  $( 2 +(2\nu+k)/s)^s = 2^s (1+(\nu+k/2)/s)^s\le 2^s \e^{\nu+k/2}$ in \eqref{mo20} yields
\begin{eqnarray}
S_\nu(A_\r,t) &\le&  \frac{\e}{\sqrt\pi} (C/\e)^{\nu+k/2} (2\nu+k)^{\nu +k/2}  \nonumber  \\\nonumber\\
&& +\frac{\e}{\sqrt{2}\, \pi} (2C)^{\nu+k/2}  \sum_{s\ge 1} (2C\varkappa^2)^s \big( s+\nu+k/2\big)^{\nu+k/2}\nonumber \\
&\le&  \frac{\e}{\sqrt\pi} (C/\e)^{\nu+k/2} (2\nu+k)^{\nu +k/2} \nonumber\\
&& + \frac{\e}{\sqrt{2}\, \pi} (2C)^{\nu+k/2} (2C\varkappa^2)^{-\nu-k/2} \sum_{s\ge \nu+k/2} (2C\varkappa^2)^s s^{\nu+k/2}.
\label{mo21}
\end{eqnarray}
To estimate the last series, we use the equality $s^{\nu+k/2} =\partial^{\nu+k/2}_\alpha|_{\alpha=0}\, \e^{\alpha s}$ to obtain
\begin{equation}
\label{mo22}
\sum_{s\ge \nu+k/2} (2C\varkappa^2)^s s^{\nu+k/2} \le \sum_{s\ge 0} (2C\varkappa^2)^s s^{\nu+k/2} =  \partial^{\nu+k/2}_\alpha|_{\alpha=0}  \,(1-2e^\alpha C\varkappa^2)^{-1},
\end{equation}
which holds provided $2C\varkappa^2<1$. Combining the bound
\begin{equation}
\label{mo23}
\big| \ \partial^m_\alpha|_{\alpha=0} \,(1-2e^\alpha C\varkappa^2)^{-1}\ \big| \le 2^m m!\max_{1\le j\le m} (1-2C\varkappa^2)^{-j} = \big(\frac{2}{1-2C\varkappa^2}\big)^m m!
\end{equation}
(for $m=\nu+k/2$) with \eqref{mo21} and \eqref{mo22}, we arrive at 
\begin{eqnarray*}
S_\nu(A_\r,t) &\le&  \frac{\e}{\sqrt\pi} (C/\e)^{\nu+k/2} (2\nu+k)^{\nu +k/2} \nonumber\\
&& +\frac{\e}{\sqrt{2}\, \pi} (2C)^{\nu+k/2} (2C\varkappa^2[1-2C\varkappa^2])^{-\nu-k/2} (\nu+k/2)!\,.
\end{eqnarray*}
Therefore, $(S_\nu(A_\r,t))^{1/\nu}$ grows at most as $((\nu+k/2)!)^{1/\nu} \sim \nu$ for large values of $\nu$. The result of Lemma \ref{lemma2} follows.\hfill $\blacksquare$

\subsubsection{Proof of Lemma \ref{lemma3}}  Since $G_j(t)=G_j$ for all times, the commutator terms $T_t$ in \eqref{mo36} simply equal 
$T_t = G^2_{n+1}\cdots G^2_{n+r/2} \, [B(t_r),[\cdots,[B(t_1), A_\r(t)]\cdots ]]$, 
and so
\begin{eqnarray}
X_0(A_\r) &=&\sum_{q\ge 1} \frac{(-\varkappa)^q}{q!}
\int_0^t dt_1\cdots\int_0^{t_{2q-1}}d t_{2q}\  \mu_\r\Big( [ B(t_{2q}), [\cdots, [B(t_1), A_\r]\cdots] ] \Big)\nonumber\\
&=& \sum_{r\ge 2\,  {\rm even}} \frac{(-\varkappa)^{r/2}}{(r/2)!}
\int_0^t dt_1\cdots\int_0^{t_{r-1}}d t_r\  \mu_\r\Big( [ B(t_r), [\cdots, [B(t_1), A_\r]\cdots] ] \Big)\qquad 
\label{mo37}
\end{eqnarray}
Next we introduce an `ancilla' harmonic oscillator with creation and annihilation operator $a^*$ and $a$, satisfying $[a,a^*]=\bbbone$, acting on the Hilbert space ${\mathcal H}_{\rm HO}$. Let $\varphi = \frac{1}{\sqrt 2}(a^*+a)$ be the harmonic oscillator field operator and consider the multicommutator acting on the Hilbert space ${\mathcal H}_{{\rm HO}}\otimes{\cal H} _\r$, $r\ge 1$,
\begin{equation}
\label{mo37.1}
 [ \varphi\otimes B(t_r), [\cdots, [ \varphi\otimes B(t_1), \bbbone_{\rm HO}\otimes A_\r]\cdots] ] = \varphi^r\otimes  [  B(t_r), [\cdots, [ B(t_1), A_\r]\cdots] ].
\end{equation}
The vacuum state $\mu_{\rm HO} = \langle\Omega_{\rm HO}, \cdot\, \Omega_{\rm HO}\rangle$ satisfies $\mu_{\rm HO}(\varphi^2)=1/2$ and by Wick's theorem,
\begin{equation}
\label{mo37.4}
\mu_{\rm HO}(\varphi^r) = 
\left\{
\begin{array}{ll}
\frac{r!}{2^r(r/2)!} & \mbox{$r$ even}\\
0 & \mbox{$r$ odd}
\end{array}
\right. .
\end{equation}
As the odd moments of $\varphi$  in $\mu_{\rm HO}$ vanish, and $\frac{1}{(r/2)!} =\frac{2^r}{r!}\mu_{\rm HO}(\varphi^r)$ for even $r$,  we obtain from \eqref{mo37} and  \eqref{mo37.1} 
\begin{eqnarray}
X_0(A_\r) &=&\sum_{r\ge 1} \frac{(2\i\sqrt{\varkappa})^r}{r!}
\int_0^t dt_1\cdots\int_0^{t_{r-1}}d t_r\nonumber\\
&& \ \ \  \mu_{\rm HO}\otimes\mu_\r\Big( [ \varphi\otimes B(t_r), [\cdots, [\varphi\otimes B(t_1),\bbbone_{\rm HO} \otimes A_\r]\cdots] ] \Big).\qquad 
\label{mo37.2}
\end{eqnarray}
The series on the right side \eqref{mo37.2} is readily identified as a Dyson series, namely,
\begin{eqnarray}
X_0(A_\r) &=& \mu_{\rm HO}\otimes\mu_\r\Big( \e^{\i t (H_\r+2\sqrt{\varkappa} \varphi \otimes B)}\e^{-\i tH_\r}  \big(\bbbone_{\rm HO}\otimes A_\r\big)\e^{\i tH_\r}\e^{-\i t (H_\r+2\sqrt{\varkappa} \varphi \otimes B)}\Big)\nonumber\\
&&  - \mu_\r (A_\r).
\label{mo37.3}
\end{eqnarray}
Combining this with \eqref{mo65} yields
\begin{equation}
\label{mo37.5}
\omega_N \big(\tau_{\lambda,N}^t(A_\r)\big) = \mu_{\rm HO}\otimes\mu_\r\Big( \e^{\i t (H_\r+2\sqrt{\varkappa} \varphi \otimes B)}  \big(\bbbone_{\rm HO}\otimes A_\r\big) \e^{-\i t (H_\r+2\sqrt{\varkappa} \varphi \otimes B)}\Big)+o_N.
\end{equation}
This proves Lemma \ref{lemma3}.\hfill $\blacksquare$

\subsubsection{Proof of Lemma \ref{fluctlem}} An application of theorem \ref{mainthmsymmetric} yields
\begin{equation}
\label{mo71}
\omega_N(F_N(A,t)) = \frac1N\sum_{n=1}^N Y_0\big(A_n(t)\big) +o(N),
\end{equation}
where $o(N)\rightarrow 0$ as $N\rightarrow\infty$ and $Y_0$ is given by \eqref{mo60}. Consider $Y_0(A_n(t))$ for a fixed $n$. For $\nu=0$, the sum over $p_1,\ldots,p_n$ in \eqref{mo60} has only terms where exactly {\em one} of the $p_j$ equals $1$ and all others vanish. Since the observable is $A_n(t)\in{\cal M}_n$, only $p_n=1$ contributes (all other terms vanish as for them, we have $T_t=0$). This forces $p_1=\cdots = p_{n-1}=0$, $p_n=1$ in \eqref{mo60}. The relation \eqref{mo75.4} then follows from \eqref{mo60} by using that the system is symmetric. \hfill $\blacksquare$

\subsubsection{Proof of Lemma \ref{leadingorderthm}}  Consider the lowest order of $X_{0,N}$ in $\lambda$ given by $r=2$ in \eqref{c4}. For this term we have $p_1=\cdots =p_n=0$ and  $p_{n+1}+\cdots +p_N=2$. Due to the vanishing odd moment condition, the last constraint implies that exactly one of the $p_j$, for a single $j\in\{n+1,\ldots, N\}$, equals $2$ and all other $p_j$ are zero. Therefore, the term with $r=0$ in \eqref{c4} equals
\begin{equation}
\label{mo24}
-\lambda^2 \sum_{j=n+1}^N \int_0^t dt_1\int_0^{t_1} d t_2  \ \omega_N\Big(  \big[G_j(t_2)\otimes B_j(t_2), [  G_j(t_1)\otimes B_j(t_1), A(t)]\big]\Big).
\end{equation}
Taking into account that $A(t)=A_\s(t)\otimes A_\r(t)$ with $A_\s(t)\in{\mathcal M}_{\le n}$ commuting with $G_j$ for $j\ge n+1$, we expand the double commutator in \eqref{mo24} to obtain
\begin{eqnarray}
\label{mo25}
\lefteqn{
	\omega_N\Big(  \big[G_j(t_2)\otimes B_j(t_2), [  G_j(t_1)\otimes B_j(t_1), A(t)]\big]\Big)}\nonumber\\
&=& 2 \mu_{\le n}(A_\s(t)) \ {\rm Re\,} \mu_j(G_j(t_2)G_j(t_1)) \, \mu_\r\big(B_j(t_2) [B_j(t_1), A_\r(t)]\big).
\end{eqnarray}
This yields the term with the double integral on the right side of \eqref{mo26}. The terms with $r>2$ in \eqref{c4} give an error $O(\lambda^4)$, uniformly in $N$.

Next take the term $r=1$ in $Y_{0,N}(A(t))$, \eqref{c5}. The constraint on the  $p_1,\ldots,p_n$  is that exactly one of them equals one, all others vanish. Moreover, $p_{n+1} = \cdots = p_N=0$. Thus the term with $r=1$ is
\begin{equation}
\label{mo26}
i\lambda \sum_{j=1}^n\int_0^t d s \ \omega_N\big( [G_j(s)\otimes B_j(s), A(t)]\big).
\end{equation}
The integrand, $\mu_{\le n}\big(G_j(s)A_\s(t)\big)\, \mu_\r\big([B_j(s),A_\r(t)]\big) + \mu_{\le n}\big([G_j(s), A_\s(t)]\big)\, \mu_\r\big(A_\r(t)B_j(s)\big)$, is readily seen to become that of the integral $\propto\lambda/\sqrt N$ on the right side of \eqref{mo27}. The remaining terms in the series \eqref{c5}, for $r>1$, give an error $O(\lambda^3)$. 

Finally, the remaining series of all terms with $\nu>0$ in \eqref{c7} adds up to an error $O(1/N)$.  \hfill $\blacksquare$

\bigskip
\noindent
{\bf Ethics statement.\ } This work did not involve any collection of human data.

\medskip
\noindent
{\bf Data accessibility statement.\ } This work does not have any experimental data.

\medskip
\noindent
{\bf  Competing interests statement.\ } We have no competing interests.

\medskip
\noindent
{\bf Acknowledgements statement.\ } We thank G.P. Berman for valuable discussions.

\medskip
\noindent
{\bf Authors' contributions statement.\ } MM conceived the mathematical model and both MM and AR worked on obtaining the results and crafting the proofs. Both authors were involved in the writing of the paper and both authors gave final approval for publication. 

\medskip
\noindent
{\bf Funding.\ } Both authors were supported by a Discovery Grant (PI MM) of the  Natural Sciences and Engineering Research Council of Canada (NSERC).

\end{document}